\documentclass[showkeys,nofootinbib,prd]{revtex4}

\usepackage{amsmath}
\usepackage{amsfonts}
\usepackage{amssymb}
\usepackage{amsthm}
\usepackage{mathtools}
\usepackage{subfigure}
\DeclareFontFamily{U}{mathb}{\hyphenchar\font45}
\DeclareFontShape{U}{mathb}{m}{n}{
      <5> <6> <7> <8> <9> <10> gen * mathb
      <10.95> mathb10 <12> <14.4> <17.28> <20.74> <24.88> mathb12
      }{}
\DeclareSymbolFont{mathb}{U}{mathb}{m}{n}

\DeclareMathSymbol{\Sun}{3}{mathb}{"40}
\allowdisplaybreaks

\begin{document}
\title{Cosmological extra dimensions can mimic dark energy}
\author{Mattia Villani}
\affiliation{University of Urbino Carlo Bo, Department of Pure and Applied Sciences (DiSPeA), Via Santa Chiara, 27, Urbino (PU), 61029, Italy}
%\affiliation{Independent scholar}
\email{mattia.villani@uniurb.it}
%\email{mattiav25@gmail.com}

\begin{abstract}
We present a simple model of a higher dimensional spacetime in which the 4d submanifold is a FLRW metric, while the extra dimensions are compactified to a hypersphere. We calculate the Friedmann equations of this model finding that the extra dimensions interact in a non-trivial way with the 4d submanifold evolution. We prove that what is observed as Dark Energy could be explained only as an effect due to the Gaussian curvature and the expansion rate of the extra dimensions. We calculate the luminosity distance in this model and derive the acceleration parameter $q$. We find that $\Omega_m$ and $q$ are in some sense \emph{dressed} by terms coming from the extra dimensions, thus a Universe filled only with matter (both luminous and dark) and with extra dimensions can explain the observational data. {We fit our model to SNe data and find evidence for the presence of 3-4 extra dimensions which are presently contracting.}
\end{abstract}
\keywords{Gravity, higher dimensional; Cosmology; Dark Energy}
\maketitle

\section{Introduction}

The most successful model that describes the present state and the past evolution of the Universe is $\Lambda$CDM model, see, for example, \cite{libro}. This model uses a background metric of the Friedmann-Lema\^{i}tre-Robertson-Walker (FLRW) type with a curvature parameter $\kappa=0$ (a {spatially} flat spacetime) and contains several types of matter/energy: one the one side there is baryonic (or visible) matter, the one that interacts with electromagnetic fields and composes nebulae,  stars, planets and all life forms; then there is Dark Matter, an unknown form of non-baryonic matter that does not interact with electromagnetic fields, but whose effects can be observed studying its gravitational effects on visible matter, see, for example, \cite{dm1,dm2}; finally there is an even more mysterious form of energy, the Dark Energy, that fills the Universe and is leading the accelerated expansion of the Universe. The most recent observations impose that the Universe is spatially flat and $\Omega_m\approx0.31$ and $\Omega_\Lambda\approx0.69$, where $\Omega_m$ is the sum of visible and Dark matter and $\Omega_\Lambda$ is the density of Dark Energy \cite{mat,mat2}.

There are several proposed theories on the nature of Dark Energy. The most simple and the one that is considered in the $\Lambda$CDM model describes the Dark Energy as a cosmological constant term $\Lambda$ in the Einstein equations \cite{lam1,lam2,lam3}; however, there is a problem with this model: the cosmological constant term calculated with Quantum Field Theory is several orders of magnitude off with respect to the measured one \cite{lam2}. {It is also possible that Dark Energy evolves with time; there are several proposals for an equation of state (EoS) for Dark Energy; probably the most used are the constant EoS (although it has some problems and might not capture all the complexity of Dark Energy behavior \cite{de1,de2,de3,de4,de5}) and the CPL parameterization \cite{de6,de7} given by}
\begin{equation}
    w(z)=w_0+w_a\,(1-a).
\end{equation}
{From the observational point of view, in the CPL case we have that $w_0=-0.725,w_a=-1.32$ \cite{cpl}.} For the case of an evolving Hubble parameter and its link to the Hubble tension, see also \cite{evo1,evo2,evo3,evo4,evo5,evo6,evo7,evo8}.

Other proposed theories suppose that Dark Energy is a scalar field with peculiar characteristics (for example, quintessence model \cite{Q1,Q2,Q3,Q4} or K-essence model \cite{k1,k2}). {CPL EoS has been used also to reconstruct the quintessence-type scalar field potential in order to describe DESI data in \cite{quint}.} 

{In other approaches, cosmologists try to modify Einstein equations by changing the Lagrangian, in particular by considering a Lagrangian of the generic form}
\begin{equation}
    \mathcal{L}=f(R)
\end{equation}
{where $R$ is the Ricci scalar and $f$ is a generic function} \cite{fr1,fr2,fr3}. See also the review \cite{review,review2}. Finally, holographic models have also been proposed \cite{holo}.

Another problem with the $\Lambda$CDM model is the so called ``Coincidence Problem'' \cite{coinc}, i.e., the fact that it seems that Dark Energy has become relevant on cosmological scale (i.e, $\Omega_\Lambda\approx\Omega_m$) only recently around $z\approx0.55$. There are several proposed solutions for this problem: one is that Dark Energy evolves with time, another emphasizes the importance of the collapse of cosmological large scale structures, in fact these structures have become non-linear in a period fairly close to the moment in which Dark Energy has become cosmologically relevant.

In parallel with the development of Cosmology, theorists have been dealing with the topic of Quantum Gravity (see, for example, \cite{QG}), a theory that could bring together General Relativity and Quantum Mechanics, solving several theoretical problems, such as that of the nature of singularities. Some of these theories, most notably String Theory (see, for example, the introductory books \cite{st1,st2}), suppose the existence of more than 4 dimensions. These extra dimensions should be compactified in such a way that the observational evidence that macroscopically there are only three spatial dimensions plus time is met.

In this paper, we introduce a simple model consisting of $4+n$-dimensional spacetime in which the 4d submanifold is described by a FLRW metric with scale factor $a$ and the $n$ extra dimensions are compactified to a hypersphere with radius $R$, in principle dependent on time and different from $a$. We were able to calculate the Friedmann equations of this model; there are three of them: the first two are modified versions of the \emph{usual} ones, and describe the evolution of the scale factor $a$, while the third describes the evolution of the radius $R$. These equations are coupled: the evolution of $a$ is tied to the evolution of $R$ in a non-trivial way. Analyzing these equations, in particular the first, which defines the Hubble parameter $H$ of the 4d submanifold, we could link the presence of the extra dimensions with Dark Energy; in particular, we shall show that $\Omega_\Lambda$ depends on the Gaussian curvature $G=R^{-2}$ and the Hubble-like parameter $\mathcal{H}=\dot{R}/R$ which describes the expansion of the extra dimensions. Moreover, as we shall see, the matter density is \emph{dressed} by the presence of the extra dimensions; in particular, the measured $\Omega_m^{Meas}$ will depend on the actual content of (visible and Dark) matter $\Omega_m^{Real}$ plus a contribution due to $G$. We shall also calculate the luminosity distance and derive the acceleration parameter, $q^{Meas}$, finding that it is also \emph{dressed} by the presence of the extra dimensions. Therefore we are able to describe the evolution of the Universe without introducing a Dark Energy component. Calculating the evolution of $\Omega_\Lambda$ term as a function of the redshift, we find that indeed, in the past its contribution was smaller; then we can also naturally solve the ``Coincidence Problem''. Finally, we study the contribution to the CMB spectrum of the extra dimensions, finding that there is none. {We refer the reader to \cite{RS1,RS2,RS3,RS4,RS5} for similar works.}

Admittedly, in our model we substitute something obscure, the Dark Energy, with something even more obscure (extra dimensions) of which, as yet, we do not have experimental evidence (there are only upper limits on their size see, for example,  \cite{exp1,exp2,exp3,exp4,exp5,bh1,ap1,ap2,ap3,ap4,eotwash}); however, we think this is an interesting approach that could be followed in future works.

This paper is organized as follows: in Section \ref{sec:metric}, we introduce our metric and calculate the Friedmann equations; in Section \ref{sec:DE}, we give an expression of $\Omega_\Lambda$ and $\Omega_m$ in terms of $G$ and $\mathcal{H}$; in Section \ref{sec:lum}, we calculate the luminosity distance and the acceleration parameter $q$; in Section \ref{sec:evol}, we describe the evolution of $\Omega_\Lambda$ with the redshift; {in Section \ref{se:comp} we fit our luminosity distance to SNe data};  finally, in Section \ref{sec:con}, we conclude our exposition.

\section{Metric and Friedmann equations}
\label{sec:metric}

We consider a metric of the form\footnote{$d\Omega_n$ is the metric of $n$-dimensional hypersphere.}
\begin{equation}\label{eq:met}
    ds^2=dt^2-\dfrac{a(t)^2}{1-r^2\,\kappa}dr^2-a(t)^2d\Omega_2-R(t)^2\,d\Omega_n
\end{equation}
i.e., a metric composed of a 4d FLRW metric with scale factor $a(t)$ and a n-dimensional submanifold compactified to a hypersphere with time-dependent radius $R(t)$, in principle different from the 4d scale factor; $\kappa=\{-1,0,1\}$ for a hyperbolic, flat and spherical 4d submanifold, respectively. The stress energy tensor is that of a perfect fluid without pressure:
\begin{equation}
    T_{00}=\rho, \qquad {T_{0i}=0} \qquad T_{ij}=0, \quad i,j=\{1,2,\dots\}.
\end{equation}

Using xAct,\footnote{\url{https://www.xact.es/}} one could derive the Einstein equations {for generic number of extra dimension $n$}, which are given by
\begin{align}\label{eq:prima}
    G_{00}&={\dfrac{\kappa}{a^2}+\left(\dfrac{\dot{a}}{a}\right)^2+n\,\dfrac{\dot{a}}{a}\dfrac{\dot{R}}{R}+\dfrac{n(n-1)}{6}\,\left[\dfrac{1}{R^2}+\left(\dfrac{\dot{R}}{R}\right)^2\right]=\dfrac{8\pi G \,\rho}{3}},\\\label{eq:seconda}
    G_{11}&={\dfrac{\kappa}{a^2}+\left(\dfrac{\dot{a}}{a}\right)^2+2\,\left( 2\dfrac{\dot{a}}{a} +\dfrac{\ddot{a}}{a} \right)+ 2n\,\dfrac{\dot{a}}{a}\,\dfrac{\dot{R}}{R}+\dfrac{n(n-1)}{2}\,\left[\dfrac{1}{R^2}+\left(\dfrac{\dot{R}}{R}\right)^2\right]+n \dfrac{\ddot{R}}{R}=0},\\\label{eq:terza}
    G_{44}&={3\dfrac{\kappa}{a^2}+3\left(\dfrac{\dot{a}}{a}\right)^2+3\dfrac{\ddot{a}}{a}+3\,(n-1)\,\dfrac{\dot{a}}{a}\,\dfrac{\dot{R}}{R}-\dfrac{(n-1)(n-2)}{2}\,\left[ \dfrac{1}{R^2}+\left( \dfrac{\dot{R}}{R} \right)^2 \right]+n\dfrac{\ddot{R}}{R}=0}.
\end{align}
We derive $\ddot{R}/R$ from the third and substitute it into the second together with $\dot{a}^2/a^2$ derived from the first, obtaining
\begin{equation}\label{eq:questa}
    \dfrac{\ddot{a}}{a}-\dfrac{n(n-1)}{3}\,\left[ \dfrac{1}{R^2}+\left( \dfrac{\dot{R}}{R} \right)^2 \right]+n\,\dfrac{\dot{a}}{a}\dfrac{\dot{R}}{R}+\dfrac{1+2n}{2+n}\,\dfrac{8\pi G\,\rho}{3}=0.
\end{equation}
We notice that equation \eqref{eq:questa} has a different coefficient in front of $\rho$ than in the \emph{usual} case, where we have $4\rho/3$, \cite{libro}: this is due to the contributions of equation \eqref{eq:terza}.

Next, we substitute \eqref{eq:prima} into \eqref{eq:terza} and compare with equation \eqref{eq:questa}, thus finding
\begin{equation}
    \dfrac{\ddot{a}}{a}-\dfrac{n}{3}\,\dfrac{\ddot{R}}{R}+\dfrac{1+n}{2+n}\,\dfrac{8\pi G\,\rho}{3}=0
\end{equation}

Therefore we have that the dynamic of our Universe is given by the two modified version of the \emph{usual} {cosmological} equations
\begin{subequations}
    \begin{equation}\label{eq:fr1}
        {\dfrac{\kappa}{a^2}+\left(\dfrac{\dot{a}}{a}\right)^2+n\,\dfrac{\dot{a}}{a}\dfrac{\dot{R}}{R}+\dfrac{n(n-1)}{3}\,\left[\dfrac{1}{R^2}+\left(\dfrac{\dot{R}}{R}\right)^2\right]=\dfrac{8\pi G \,\rho}{3},}
    \end{equation}
    \begin{equation}\label{eq:fr2}
        {\dfrac{\ddot{a}}{a}-\dfrac{n(n-1)}{3}\,\left[ \dfrac{1}{R^2}+\left( \dfrac{\dot{R}}{R} \right)^2 \right]+n\,\dfrac{\dot{a}}{a}\dfrac{\dot{R}}{R}+\dfrac{1+2n}{2+n}\,\dfrac{8\pi G\,\rho}{3}=0,}
    \end{equation}
which describe the evolution of $a$, together with the new equation
    \begin{equation}\label{eq:fr3}
    \begin{split}
        \dfrac{\ddot{R}}{R}&=\dfrac{3}{n}\dfrac{\ddot{a}}{a}+\dfrac{1+n}{2+n}\,\dfrac{8\pi G\,\rho}{n}\\
        &{=(1-n)\,\left[ \dfrac{1}{R^2}+\left( \dfrac{\dot{R}}{R} \right)^2 \right]-3\,\dfrac{\dot{a}}{a}\dfrac{\dot{R}}{R}+\dfrac{8\pi G\,\rho}{(2+n)},}
    \end{split}
    \end{equation}
which describes the evolution of the extra dimensions. We notice that the radius $R$ and its derivatives enter in a not-trivial way into the two modified Friedmann equations, thus extra dimensions can affect the dynamic of the scale factor in a complicated way. We notice, furthermore, that we have considered a Universe filled only with dust, but the equations could be readily extended including a fluid with pressure. {Finally, we notice that the first line of equation \eqref{eq:fr3} and equation \eqref{eq:fr2} are consistent with each other in the limit $n\rightarrow0$ (in order to see this, one has to multiply \eqref{eq:fr3} by $n/3$) and then take the limit.} We were not able to find a closed form or parametric solutions to the given equations.\footnote{Actually, for the flat case $\kappa=0$ we were able to give a solution in form of a series expansion of $a$ and $R$, but it is too complicated to be of any practical use.}
\end{subequations}

\section{Extra dimensions as dark energy and effects on matter density}
\label{sec:DE}

If we solve equation \eqref{eq:fr1} for $\dot{a}/a$, we obtain
\begin{equation}\label{eq:hubble}
    {\left( \dfrac{\dot{a}}{a} \right)^2\,\dfrac{1}{H_0^2}=\dfrac{\Omega_m}{a^3}-\dfrac{\Omega_k}{a^2}-\dfrac{n(n-1)}{6}\,\dfrac{\Omega_G}{H_0^2}+\dfrac{n(1+2n)}{6}\dfrac{\mathcal{H}^2}{H_0^2}-\dfrac{n\,\mathcal{H}}{H_0}\,\sqrt{-\dfrac{n(n-1)}{6}\,\dfrac{\Omega_G}{H_0^2}+\dfrac{\Omega_m}{a^3}-\dfrac{\Omega_k}{a^2}+\dfrac{n(n+2)}{12}\,\dfrac{\mathcal{H}^2}{H_0^2}}}
\end{equation}
where $\Omega_m$, $\Omega_k$ and $H_0$ have the usual meaning and $\Omega_G=R^{-2}$ is the Gaussian curvature of the extra dimensions and $\mathcal{H}=\dot{R}/R$, is a sort of Hubble parameter for the extra dimensions with the same dimensions of $H_0$ (km s$^{-1}$ Mpc$^{-1}$. If we evaluate \eqref{eq:hubble} at present time, we get
\begin{equation}\label{eq:quella}
    {1=\Omega_m-\Omega_k-\dfrac{n(n-1)}{6}\,\dfrac{\Omega_G}{H_0^2}+\dfrac{n(1+2n)}{6}\,\dfrac{\mathcal{H}_0^2}{H_0^2}-\dfrac{n\,\mathcal{H}_0}{H_0}\,\sqrt{\Omega_m-\Omega_k-\dfrac{n(n-1)}{6}\,\dfrac{\Omega_G}{H_0}+\dfrac{n(n+2)}{12}\,\dfrac{\mathcal{H}_0^2}{H_0^2}}.}
\end{equation}
If we assume that the 4d FLRW metric is flat $\Omega_k=0$ (a geometry that is favored observationally \cite{mat2}), then, since we also assume that there is no Dark Energy, $\Omega_m=\Omega_m^{Real}=1$, but we can still have the measured value $\Omega_m^{Meas}\approx0.31$ if we impose ({compare to \cite{RS4}})
\begin{equation}\label{eq:matter}
    \Omega_m^{Meas}=\Omega_m^{Real}-\dfrac{n(n-1)}{6}\,\dfrac{\Omega_G}{H_0^2}.
\end{equation}
Thus we see that the matter density is in some sense \emph{dressed} by the curvature of the extra dimensions, which behaves as a matter field with negative mass (see ref for a similar result). We can then continue with our analogy and impose
\begin{equation}
\begin{split}
    \Omega_\Lambda^0&=\dfrac{n(1+2n)}{6}\,\dfrac{\mathcal{H}_0^2}{H_0^2}-\dfrac{n\,\mathcal{H}_0}{H_0}\,\sqrt{\Omega^{Real}_m-\dfrac{n(n-1)}{6}\,\dfrac{\Omega_G}{H_0^2}+\dfrac{n(n+2)}{12}\,\dfrac{\mathcal{H}_0^2}{H_0^2}}\\
    &=\dfrac{n(1+2n)}{6}\,\dfrac{\mathcal{H}_0^2}{H_0^2}-\dfrac{n\,\mathcal{H}_0}{H_0}\,\sqrt{\Omega^{Meas}_m+\dfrac{n(n+2)}{12}\,\dfrac{\mathcal{H}_0^2}{H_0^2}},
\end{split}
\end{equation}
i.e., the term that in the $\Lambda$CDM model is due to Dark Energy in our model is given by a combination of the expansion rate and of the Gaussian curvature of the extra dimensions.

%Substituting the measured values of $H_0$, $\Omega_m^{Meas}$ and $\Omega_\Lambda^0$ \cite{mat,mat2,tension}, we can obtain the present values of the unknown parameters
%\begin{equation}
%    G=2.07\,H_0^2, \qquad \mathcal{H}_0=-0.366\,H_0.
%\end{equation}
%We then see that we can completely eliminate the need of introducing Dark Energy into our model using a FLRW metric with extra dimensions which are contracting, since the above equations mean $\left.\dot{R}\right|_0<0$.

In the following Section we derive the expression of the luminosity distance.

\section{Luminosity distance}
\label{sec:lum}

The luminosity distance is given by
\begin{equation}
    L(z)=(1+z)\, \int_0^z \dfrac{dz^\prime\,c}{H(z^\prime)}
\end{equation}
where $H(z)$ is given by equation \eqref{eq:hubble}. We need to expand it for small redshift. At first order, we have, by definition (see, for example, \cite{capo})
\begin{equation}
    \dfrac{dH}{dz}=H\,(1+q^{Meas}), \qquad q^{Meas}=-\dfrac{1}{a}\dfrac{\ddot{a}}{H^2}
\end{equation}
where $q^{Meas}$ is the measured acceleration parameter. We can derive an expression for this parameter by explicitly deriving \eqref{eq:hubble} and substituting\footnote{In order to derive these expression, one should remember that
\begin{equation*}
    \dfrac{d}{dt}=-(1+z)\,H\,\dfrac{d}{dz}.
\end{equation*}}
\begin{equation}
    \dfrac{dR}{dz}=-R\,\dfrac{\mathcal{H}}{H_0}\, \qquad \dfrac{d\mathcal{H}}{dz}=-\dfrac{\mathcal{H}^2}{H_0}(-1+q_e)\, \qquad q_e=\dfrac{\ddot{R}}{R}\,\dfrac{1}{\mathcal{H}^2}
\end{equation}
where $q_e$ is the acceleration parameter of the extra dimensions (pay attention to the sign convention we are using). {We notice that from the first line in equation \eqref{eq:fr3}, we obtain}
\begin{equation}\label{eq:acc}
    {q_e=\dfrac{3}{n}\,\left( -q+\dfrac{n+1}{n+2}\,\Omega_m^{Real} \right)\,\dfrac{\mathcal{H}_0}{H_0}.}
\end{equation}
Using the relation 
\begin{equation}
\kappa=-1+\Omega_m  -\dfrac{n(n-1)}{6}\,\dfrac{\Omega_G}{H_0^2}-n\,\dfrac{\mathcal{H}_0}{H_0}-\dfrac{n(n-1)}{6}\,\dfrac{\mathcal{H}_0^2}{H_0^2}  
\end{equation}
obtained from equation \eqref{eq:fr1} {and the definition of the acceleration parameter $q_e$ from equation \eqref{eq:acc}, we obtain (using also $q=\Omega_m^{Real}/2$)}
\begin{equation}\label{eq:qmeas}
\begin{split}
    q^{Meas}&={\dfrac{\Omega_m}{2}\,\left( 1-\dfrac{3}{2+n}+\dfrac{3H_0}{2H_0+\mathcal{H}_0n} \right)-\dfrac{n\,(\mathcal{H}_0-H_0)\,\left[ \mathcal{H}_0\,(3\,H_0+\mathcal{H}_0\,(n-1))+\Omega_G\,(n-1) \right]}{3\,H_0^2\,(2H_0+\mathcal{H}_0\,n)}}=\\
    &={\dfrac{\Omega_m}{2}\,\left( 1-\dfrac{3}{2+n}+\dfrac{3}{2+h_0\,n} \right)-\dfrac{n\,(h_0-1)\,\left[ h_0\,(3+h_0\,(n-1))+\Omega^{red}_G\,(n-1) \right]}{3\,(2+h_0\,n)}},
\end{split}    
\end{equation}
{where in the second line we have defined the reduced parameters $\mathcal{H}_0=h_0\,H_0$, $\Omega_G=\Omega_G^{red}\,H_0^2$, where $h_0$ and $\Omega_G^{red}$ are pure numbers. Thus, also the real acceleration parameter $q^{Real}=\Omega_m^{Real}/2$ is \emph{dressed} by terms depending on the dynamic of the extra dimensions and together contribute to its measured value, $q^{Meas}$. One can check that in the limit $n\rightarrow0$, the right hand side of the above equation reduces to $q^{Real}=q^{Meas}=\Omega_m^{Real}/2$, as it should. }

%If we consider as above a flat 4d spacetime and substitute the measured value of $q^{Meas}$ \cite{mat,mat2,tension}, the values of $G$ and $\mathcal{H}_0$ obtained above and the real value of the acceleration parameter $q^{Real}=\Omega_m^{Real}/2$, we can derive $q_e$:
%\begin{equation}
%    q_e=-8.83
%\end{equation}
%Thus, to explain the observations, the contraction of the extra dimensions must decelerate. We notice how large this acceleration parameter is compared to that of the 4d submanifold.

The luminosity distance up to $O(z^3)$ is given by
\begin{equation}\label{eq:lum}
    L(z)=\dfrac{cz}{H_0}+\dfrac{cz^2}{2H_0}\,\left( 1-\dfrac{\Omega_m}{2}\,\left( 1-\dfrac{3}{2+n}+\dfrac{3}{2+h_0\,n} \right)+\dfrac{n\,(h_0-1)\,\left[ h_0\,(3+h_0\,(n-1))+\Omega^{red}_G\,(n-1) \right]}{3\,(2+h_0\,n)} \right)+O(z^3),
\end{equation}
{which reduces to the usual one with $q=\Omega_m/2$ when $n\rightarrow0$.}

\section{The evolution of the Dark Energy term}
\label{sec:evol}

In the above Section, we have found that the present value of the Dark energy density is
\begin{equation*}
    \Omega_\Lambda^0=\dfrac{n(1+2n)}{6}\,\dfrac{\mathcal{H}_0^2}{H_0^2}-\dfrac{n\,\mathcal{H}_0}{H_0}\,\sqrt{\Omega^{Meas}_m+\dfrac{n(n+2)}{12}\,\dfrac{\mathcal{H}_0^2}{H_0^2}}.
\end{equation*}
If we restore the redshift dependence and the effects of curvature, we have
\begin{equation}
    \Omega_{\Lambda}(z)=\dfrac{n(1+2n)}{6}\,\dfrac{\mathcal{H}^2(z)}{H_0^2}-\dfrac{n\,\mathcal{H}(z)}{H_0}\,\sqrt{-\dfrac{n(n-1)}{6}\,\dfrac{\Omega_G(z)}{H_0^2}+\Omega_m\,(1+z)^3-\Omega_k\,(1+z)^2+\dfrac{n(n+2)}{12}\,\dfrac{\mathcal{H}^2(z)}{H_0^2}}.
\end{equation}
We now expand the above expression for small redshift, obtaining
\begin{equation}\label{eq:lam}
\begin{split}
    \Omega_\Lambda(z)&=\Omega_\Lambda^0+\Bigg( \dfrac{\Omega_G^{red}\,(h_0-1)\,h_0\,n^2\,(n-1)}{3\,(n\,h_0-2)}+\dfrac{n\,h_0\,(-6-h_0\,(n-1)\,(-6+h_0\,(n\,h_0+n-4)))}{3(n\,h_0-2)}+\\
    &+\dfrac{n\,\Omega_m^{Real}\,(6+h_0\,(-6+h_0\,(n-1)\,n)}{2\,(n+2)\,(n\,h_0-2)} \Bigg)\,z+O(z^2).
\end{split}
\end{equation}
We notice that in the limit $n\rightarrow0$, the above relation reduces to $\Omega_\Lambda(z)=\Omega_\Lambda^0$, as it should. %Substituting the numbers from the previous Sections, we get
%\begin{equation}
%    \Omega_\Lambda(z)=0.69-3.65\,z+O(z^2);
%\end{equation}
%this means that in the past (i.e., at large redshift) the density of the Dark Energy was smaller, and consequently the acceleration of the expansion of the Universe was also smaller than today. 

\section{Comparison with Supernovae data}
\label{se:comp}
{In this Section, we derive numerical estimate of the parameters comparing the luminosity distance we have derived with SNe data from the compilation Union 3.1 \cite{sn}. We follow \cite{sn,sn2} and fit the function (see \cite{sn} for the notation)}
\begin{equation}
    m^{pred}=-\alpha\,x_1+\beta_B\,c_B+\left[ P^{high}_{eff}\,\beta_{R,high}+(1-P^{high}_{eff})\,\beta_{R,low} \right]\,c_R-\delta(0)\,P^{high}_{eff}+M_B+\mu
\end{equation}
{where $P^{high}_{eff}$ is reported in \cite{sn3} and where}
\begin{equation}
    \mu=m-M=5\,\log\left( \dfrac{L}{10\,pc} \right)
\end{equation}
{where $L$ is the luminosity distance given in equation \eqref{eq:lum}. We fix the parameters $\alpha$, $\beta_B$, $\beta_{R,high}$, $\beta_{R,low}$, $M$ and $\delta(0)$ to the values reported in \cite{sn}, while all the other parameters are SN-dependent. The results of our fit are presented in Table \ref{tab:param}, where we report the full parameters $\mathcal{H}_0$ and $\Omega_G$, not their reduced form $h_0$ and $\Omega_G^{red}$. We notice that our fit seems to suggest within 1$\sigma$ the presence of 3-4 extra dimensions, the case of zero extra dimension being excluded to 6$\sigma$. The Hubble-like parameter $\mathcal{H}_0$ is negative, suggesting that the extra dimensions are currently contracting. The Hubble parameter $H_0$ we have found is somewhat lower than that usually found using SNe data \cite{sn,sn2,tension}, but still larger than the CMB results \cite{planck}, so we do  not solve the Hubble tension (but more work is needed in order to compare CMB data with our model); moreover, our fit seems to suggest that the 4d FLRW Universe is closed, but within 1.3$\sigma$ our model is consistent with a spatially flat Universe. $\Omega_m^{Real}$ and $\Omega_\Lambda^0$ are also different from the values found in other works \cite{sn,sn2,planck, mat,mat2}, but still compatible with them within 1$\sigma$.}

\begin{table}[ht]
    \centering
    \begin{tabular}{ccccccc}
         $H_0$ (km s$^{-1}$ Mpc$^{-1}$) & $\Omega_m^{Real}$  & $\Omega_k$& $\Omega_\Lambda^0$ & $\Omega_G$ (km$^2$ s$^{-2}$ Mpc$^{-2}$) & $\mathcal{H}_0$ (km s$^{-1}$ Mpc$^{-1}$) & $n$ \\
         \hline
         $71.9\pm0.9$ & $0.38\pm0.07$ & $0.30\pm0.23$ & $0.52\pm0.14$ & $10.35\pm0.22$ & $-20.7\pm2.7$ & $3.6\pm0.6$ 
    \end{tabular}
    \caption{Parameters and respective errors derived from a fit to SNe data.}
    \label{tab:param}
\end{table}

{With the parameter values found above, we obtain the following expression for the evolution of the cosmological constant using equation \eqref{eq:lam}}
\begin{equation}
    \Omega_\Lambda=0.52-1.64\,z+O(z^2),
\end{equation}
{thus, the value of the density of the Dark Energy was smaller in the past; this could be an explanation of the Coincidence Problem: the evolution of the extra dimensions has made the Dark Energy term grow with time, bringing it to the modern value.}

{Finally, we notice, however, that since $\Omega_m^{Real}$ is not much different from the usual value, one could also consider a \emph{dressed} curvature parameter}
\begin{equation}
    \Omega_\kappa^{Meas}=\Omega_\kappa^{Real}+\dfrac{n(n-1)}{6}\,\dfrac{\Omega_G}{H_0^2}
\end{equation}
{in place of the \emph{dressed} matter density. Due to the smallness of $\Omega_G$ in our fit, there is not much difference from the dressed and real parameters.}

\section{Conclusion}
\label{sec:con}
We have presented a simple model consisting in a $4+n$-dimensional spacetime where the 4d submanifold is given by a FLRW metric, while the n-dimensional one is given by a hypersphere with radius $R$ dependent on time. We have derived the modified Friedmann equations for this spacetime, finding that the extra dimensions affect the evolution of the 4d submanifold in a non-trivial way. We have calculated the Hubble parameter of this spacetime and found that the Gaussian curvature affects the measurement of the matter density of the Universe, while the remaining terms, containing a function describing the velocity of the expansion of the extra dimensions, behave as Dark Energy, in fact leading the acceleration of the expansion of the 4d submanifold. We have continued calculating the luminosity distance in this spacetime finding that also the acceleration parameter is \emph{dressed} by terms coming from the dynamic of the extra dimensions. {We have presented a fit to observation data of our luminosity distance; in particular, we have found that in order to explain the observations, one should have 3-4 extra dimensions contracting at the present time}. 

{Finally, we point out a problem of this model: we essentially double the number of unknown parameters; in fact, in a fit of observational data, we need 6 parameters, i.e. $H_0$, $\Omega_m^{Real}$, $\Omega_\kappa$, $\Omega_G$, $\mathcal{H}_0$, $n$, while $\Lambda$CDM model would only need 4: $H_0$, $\Omega_m^{Real}$, and $\Omega_\kappa$ or $\Omega_\Lambda$.}

\section*{Funding details}
This work did not receive any funding.

\bibliography{biblio}

\end{document}